\documentclass[reqno]{amsart}
\usepackage{hyperref}

\newtheorem{theorem}{Theorem}[section]
\newtheorem{lemma}[theorem]{Lemma}

\newtheorem{hypothesis}[theorem]{Hypothesis {\bf H.}\hspace*{-0.6ex}}

\theoremstyle{definition}

\allowdisplaybreaks

\newcommand{\R}{{\mathbb R}}
\newcommand{\N}{{\mathbb N}}
\newcommand{\Z}{{\mathbb Z}}
\newcommand{\C}{{\mathbb C}}

\newcommand{\nn}{\nonumber}
\newcommand{\be}{\begin{equation}}
\newcommand{\ee}{\end{equation}}
\newcommand{\bea}{\begin{eqnarray}}
\newcommand{\eea}{\end{eqnarray}}
\newcommand{\ul}{\underline}

\newcommand{\ol}{\overline}

\newcommand{\E}{\mathrm{e}}

\newcommand{\dott}{\,\cdot\,}

\newcommand{\f}{\frac}

\newcommand{\I}{\mathrm{i}}

\DeclareMathOperator{\AL}{AL}

\newcommand{\lb}{\label}

\numberwithin{equation}{section}

\begin{document}

\title[On asymptotics of solutions of the AL hierarchy]
{On the spatial asymptotics of solutions of the Ablowitz--Ladik hierarchy
}

\author[J. Michor]{Johanna Michor}
\address{Faculty of Mathematics\\
University of Vienna\\
Nordbergstrasse 15\\ 1090 Wien\\ Austria\\ and International Erwin Schr\"odinger
Institute for Mathematical Physics, Boltzmanngasse 9\\ 1090 Wien\\ Austria}
\email{\href{mailto:Johanna.Michor@univie.ac.at}{Johanna.Michor@univie.ac.at}}
\urladdr{\href{http://www.mat.univie.ac.at/~jmichor/}{http://www.mat.univie.ac.at/\~{}jmichor/}}

\dedicatory{Dedicated with great pleasure to Peter W. Michor on the occasion of his 60th birthday.}

\thanks{Research supported by the Austrian Science Fund (FWF) under Grant No.\ V120}

\begin{abstract}
We show that for decaying solutions of the Ablowitz--Ladik system, the 
leading asymptotic term is time independent. In addition, two arbitrary 
bounded solutions of the Ablowitz--Ladik system which are asymptotically 
close at the initial time stay close. All results are
also derived for the associated hierarchy.
\end{abstract}

\maketitle

\section{Introduction}

When solving completely integrable wave equations via the inverse scattering transform, a method
developed by Gardner et al.\ \cite{ggkm} in 1967 for the Korteweg--de Vries (KdV) equation,
one intends to prove existence of solutions within the respective class. In particular, short-range 
perturbations of the background solution should remain short-range during the time evolution.
So to what extend are spatial asymptotical properties time independent? 

For the KdV equation, this question was answered by Bondareva and Shubin~\cite{Bondareva:1983}, \cite{BondarevaShubin:1982},
who considered the Cauchy problem for initial conditions which have a prescribed asymptotic expansion 
in terms of powers of the spatial variable and showed that the leading term of the expansion
is time independent. Teschl~\cite{Teschl2009} considered the initial value problem for the Toda 
lattice in the class of decaying solutions and obtained time independence of the leading term.

In this note we want to address the same question for the Ablowitz--Ladik ($\AL$) system,
an integrable discretization of the AKNS-ZS system derived by Ablowitz and Ladik (\cite{AblowitzLadik:1975}--\cite{AblowitzLadik2:1976}) in the mid seventies. The $\AL$ system is given by 
\begin{align} \label{ALequ}
\begin{split}
-i \alpha_{t}- (1-\alpha\beta) (\alpha^- + \alpha^+) + 2 \alpha=0, \\   
 -i\beta_{t}+ (1-\alpha\beta) (\beta^- + \beta^+) - 2 \beta=0,
\end{split}
\end{align}
where $\alpha=\alpha(n,t)$, $\beta=\beta(n,t)$, $(n,t) \in \Z \times \R$, are complex valued sequences and
$f^\pm(n,t)=f(n\pm1,t)$. In the defocusing ($\beta=\overline{\alpha}$) and focusing case 
($\beta=- \overline{\alpha}$), \eqref{ALequ} is a discrete analog of the nonlinear Schr\"odinger (NLS) equation
\[ 
\I q_t + q_{xx} \pm 2 q |q|^2 = 0.
\]
We refer to the monographs \cite{AblowitzClarkson:1991}, \cite{AblowitzPrinariTrubatch:2004}, or 
\cite{GesztesyHoldenMichorTeschl:2008} for further information.

Our main result in Theorem~\ref{thm2.4} yields that the dominant term of suitably decaying 
solutions $\alpha(n,t)$, $\beta(n,t)$ of \eqref{ALequ}, for instance weighted $\ell^{2p}$ sequences 
whose spatial difference is in $\ell^p$, $1\leq p < \infty$, is time independent. For example, 
\be \label{example}
\alpha(n,t)= \frac{a}{n^\delta} + O\Big(\frac{1}{n^{\min(2\delta, \delta +1)}}\Big), \quad 
\beta(n,t)= \frac{b}{n^\delta} + O\Big(\frac{1}{n^{\min(2\delta, \delta +1)}}\Big), 
\quad n\rightarrow \infty,
\ee
holds for fixed $t$, provided it holds at the initial time $t=t_0$.
Here $a,b \in \C$ and $\delta \geq 0$.
A similar expression is valid for $n \rightarrow - \infty$.

The inverse scattering transform for the $\AL$ system with vanishing boundary conditions 
was studied in \cite{AblowitzLadik:1976}.
Ablowitz, Biondini, and Prinari~\cite{AblowitzBiondiniPrinari:2007} 
(compare also Vekslerchik and Konotop~\cite{VekslerchikKonotop:1992}) considered
nonvanishing steplike boundary conditions 
$\alpha(n) \rightarrow \alpha_0 e^{\I \theta_\pm}$ as $|n| \rightarrow \infty$, $\alpha_0 >0$,
in the class 
\be \label{steplike}
\sum_{j=n}^{\pm \infty} (\alpha(j) - \alpha_0 e^{\I \theta_\pm}) < \infty
\ee
for the defocusing discrete NLS equation. Quasi-periodic boundary conditions for the $\AL$ hierarchy 
will be considered in Michor \cite{Michor2009}.
As mentioned, a crucial step is to show that short-range perturbations like \eqref{steplike} of solutions 
stay short-range. Here we show in general that arbitrary bounded solutions of the $\AL$ system which are 
asymptotically close at the initial time stay close. 

In Section~\ref{secivp} we derive our results for the $\AL$ system and extend them 
in Section~\ref{secth} to the $\AL$ hierarchy, a completely integrable hierarchy of 
nonlinear evolution equations whose first nonlinear member is \eqref{ALequ}.

\section{The initial value problem for the Ablowitz--Ladik system}
\label{secivp}

Let us begin by recalling some basic facts on the system~\eqref{ALequ}.
We will only consider bounded solutions and hence require

\begin{hypothesis} \lb{hALh5.1}
Suppose that $\alpha, \beta \colon \Z\times\R \to \C$ satisfy
\begin{align}
\begin{split}
& \sup_{(n,t)\in\Z\times\R}\big(|\alpha(n,t)|
+|\beta(n,t)|\big) < \infty,  \lb{ALh5.1} \\
& \alpha(n,\dott), \, \beta(n,\dott) \in C^1(\R), \; n\in\Z,  \quad 
 \alpha(n,t)\beta(n,t)\notin\{0,1\}, \; (n,t)\in\Z\times \R.  
 \end{split}
\end{align}
\end{hypothesis}
The $\AL$ system~\eqref{ALequ}
is equivalent to the zero-curvature equation
\be \label{zc}
U_t+UV -V^+ U=0, 
\ee
where
\be \label{UV}
U(z)=\begin{pmatrix} z & \alpha \\ \beta z & 1 \end{pmatrix},
\quad
V(z)=i\begin{pmatrix}
z-1-\alpha\beta^- & \alpha - \alpha^- z^{-1}\\ \beta^- z -\beta &
1+ \alpha^- \beta -z^{-1} \end{pmatrix}  
\ee
for the spectral parameter $z\in\C\setminus\{0\}$. The $\AL$ system can also be formulated in terms of Lax pairs, 
see \cite{GesztesyHoldenMichorTeschl:2007b}. Then \eqref{ALequ} is equivalent to the Lax equation 
\begin{equation}\label{Lax1}
\f{d}{dt} L(t) - [P(t), L(t)] =0, \qquad t\in\R,
\end{equation}
where $L$ reads in the standard basis of $\ell^2(\Z)$ (abbreviate $\rho=(1-\alpha\beta)^{1/2}$)
\begin{align} \label{L}
L &= \left(\begin{smallmatrix} \ddots &&\hspace*{-8mm}\ddots
&\hspace*{-10mm}\ddots &\hspace*{-12mm}\ddots
&\hspace*{-14mm}\ddots &&&\raisebox{-3mm}[0mm][0mm]{\hspace*{-6mm}{\Huge $0$}}\\
&0& -\alpha(0) \rho(-1) & -\beta(-1)\alpha(0) & -\alpha(1)\rho(0) & \rho(0) \rho(1)\\
&& \rho(-1) \rho(0) & \beta(-1) \rho(0) &-\beta(0) \alpha(1) & \beta(0) \rho(1) & 0\\
&&&0& -\alpha(2) \rho(1) & -\beta(1) \alpha(2) &-\alpha(3) \rho(2) & \rho(2) \rho(3)\\
&&\raisebox{-4mm}[0mm][0mm]{\hspace*{-6mm}{\Huge $0$}} &&
\rho(1) \rho(2) & \beta(1) \rho(2) & -\beta(2) \alpha(3)& \beta(2) \rho(3) & 0\\
&&&&&\hspace*{-14mm}\ddots &\hspace*{-14mm}\ddots
&\hspace*{-14mm}\ddots &\hspace*{-8mm}\ddots &\ddots
\end{smallmatrix}\right) 
\end{align}
and $P$ is given by
\[
P = \tfrac{i}{2}\big( L_+ - L_- + (L^{-1})_- - (L^{-1})_+ +2 Q_d\big).  
\]
Here $Q_d$ is the doubly infinite diagonal matrix
$Q_d=\big((-1)^k \delta_{k,\ell} \big)_{k,\ell \in\Z}$ and $L_\pm$ denote the upper and lower
triangular parts of $L$,
\be \label{L_pm}
L_\pm=\big(L_\pm (m,n)\big)_{(m,n)\in\Z^2}, \quad
L_\pm (m,n)=\begin{cases} L(m,n), &\pm(n-m)>0, \\ 0, & \text{otherwise.}
\end{cases}
\ee

The Lax equation \eqref{Lax1} implies existence of a propagator
$W(s,t)$ such that the family of operators $L(t)$, $t \in \R$, is similar,
\[
L(s) = W(s,t)L(t) W(s,t)^{-1}, \quad s,t \in \R.
\]

By \cite[Sec.\ 3.8]{GesztesyHoldenMichorTeschl:2008} or \cite{GesztesyHoldenMichorTeschl:2007b},
existence, uniqueness, and smoothness of local solutions of the $\AL$ initial value problem 
follow from \cite[Thm 4.1.5]{AbrahamMarsdenRatiu1988}, since the $\AL$ flows are autonomous.

\begin{theorem} \lb{thm2.2} 
Let $t_{0}\in\R$ and suppose $(\alpha_0, \beta_0) \in M= \ell^p(\Z)\oplus\ell^p(\Z)$ for some 
$p\in [1,\infty)\cup\{\infty\}$. Then there exists $T>0$ and a unique local integral curve
$t\mapsto(\alpha(t), \beta(t))$ in $C^\infty((t_0-T,t_0+T),M)$ of the Ablowitz--Ladik system~\eqref{ALequ}
such that $ (\alpha,\beta)\big|_{t=t_0} 
= (\alpha_0, \beta_0)$.
\end{theorem}

Our first lemma shows that the leading asymptotics as $n \rightarrow \pm \infty$ are preserved by 
the $\AL$ flow. We only state the result for the $\AL$ system, whose proof follows as the one of Lemma~\ref{lemma1}.
Define
\be \label{norm2}
\|(\alpha, \beta)\|_{w,p} = \begin{cases}
\bigg(\sum\limits_{n\in \Z} w(n) \big( |\alpha(n)|^p+|\beta(n)|^p\big)\bigg)^{1/p}, & \quad 1\leq p<\infty, \\
\ \sup\limits_{n\in \Z}w(n)\big(|\alpha(n)|+ |\beta(n)|\big), & \quad p=\infty.
\end{cases}
\ee

\begin{lemma} \label{lemma2}
Let $w(n)\geq 1$ be some weight with $\sup_n(|\frac{w(n+1)}{w(n)}|+|\frac{w(n)}{w(n+1)}|)<\infty$.
Fix $1\leq p\leq \infty$ and suppose $(\alpha(n,t), \beta(n,t))$ and $(\tilde \alpha(n,t), \tilde \beta(n,t))$ 
are arbitrary bounded solutions of the $\AL$ system~\eqref{ALequ}. If
\be \label{eq 2.2}
\|(\alpha(t)-\tilde \alpha(t),\beta(t)-\tilde \beta(t))\|_{w,p}
< \infty
\ee
holds for one $t=t_0 \in \R$, then it holds for all $t \in (t_0-T,t_0+T)$.
\end{lemma}

But even the leading term is preserved by the time evolution. 

\begin{theorem} \label{thm2.4}
Let  $w(n)\geq 1$ be some weight with $\sup_n(|\frac{w(n+1)}{w(n)}|+|\frac{w(n)}{w(n+1)}|)<\infty$.
Fix $1\leq p\leq \infty$ and suppose $\alpha_0$, $\beta_0$ and $\tilde \alpha_0$, $\tilde \beta_0$ are bounded sequences such that
\begin{align*} \label{decay}
& \begin{array}{ll}
 \|(\alpha_0, \beta_0)\|_{w,2p} < \infty, &  \|(\alpha_0-\alpha_0^+, \beta_0-\beta_0^+)\|_{w,p} < \infty, \\
 \|(\tilde \alpha_0, \tilde \beta_0)\|_{w,p} < \infty, & \\
\end{array}\qquad \text{if $1 \leq p < \infty$,} \\[2mm]
& \begin{array}{ll}
 \|(\alpha_0, \beta_0)\|_{w,\infty} < \infty, &
\|(\alpha_0-\alpha_0^+, \beta_0-\beta_0^+)\|_{w^2,\infty} < \infty, \\
 \|(\tilde \alpha_0, \tilde \beta_0)\|_{w^2,\infty} < \infty, & 
\end{array} \qquad \text{if $p=\infty$.}
\end{align*}
Let $(\alpha(t),\beta(t))$, $t \in (-T,T)$, be the unique solution of the 
Ablowitz--Ladik system~\eqref{ALequ} corresponding to the initial conditions
\be
\alpha(0) = \alpha_0 + \tilde \alpha_0, \qquad \beta(0) = \beta_0 + \tilde \beta_0.
\ee
Then this solution is of the form
\be
\alpha(t) = \alpha_0 + \tilde \alpha(t), \qquad \beta(t) = \beta_0 + \tilde \beta(t),
\ee
where $\|(\tilde \alpha(t), \tilde \beta(t))\|_{w,p} < \infty$, respectively, 
$\|(\tilde \alpha(t), \tilde \beta(t))\|_{w^2,\infty} < \infty$.
\end{theorem}

\begin{proof}
The proof relies on the idea to consider our differential equation in two nested spaces of sequences,
the Banach space of all ${(\alpha(n), \beta(n))}$ with sup norm, and the Banach space with 
norm $\| .\|_{w,p}$, as follows.  
Plugging $(\alpha_0 + \tilde \alpha(t), \beta_0 + \tilde \beta(t))$ into the 
$\AL$ equations \eqref{ALequ} yields a differential equation for $(\tilde \alpha(t), \tilde \beta(t))$
\begin{align} \label{2.15}  \nn 
\I \tilde \alpha_t(t) &= - \big(1-(\alpha_0 + \tilde \alpha(t))(\beta_0 + \tilde \beta(t))\big) 
 \big(\tilde \alpha^+(t) + \alpha_0^+ + \tilde \alpha^-(t) + \alpha_0^- \big) + 2\tilde \alpha(t) + 2\alpha_0 \\ \nn
&= \alpha_0 - \alpha_0^- + \alpha_0 - \alpha_0^+ + \alpha_0\beta_0(\alpha_0^++\alpha_0^-) 
\\ \nn
&\quad + \tilde \alpha(t)\big(2 + (\beta_0 + \tilde \beta(t))(\tilde \alpha^+(t) + \alpha_0^+ + \tilde \alpha^-(t) + \alpha_0^-) \big)\\ \nn
&\quad + \tilde \beta(t)\alpha_0\big(\tilde \alpha^+(t) + \alpha_0^+ + \tilde \alpha^-(t) + \alpha_0^- \big)\\ \nn
&\quad+ \tilde \alpha^+(t)(\alpha_0\beta_0-1) +\tilde \alpha^-(t)(\alpha_0\beta_0-1),\\ \nn
\I \tilde \beta_t(t) &= \big(1-(\alpha_0 + \tilde \alpha(t))(\beta_0 + \tilde \beta(t)) \big) 
 \big(\tilde \beta^+(t) + \beta_0^+ + \tilde \beta^-(t) + \beta_0^- \big) - 2\tilde \beta(t) 
- 2\beta_0 \\ \nn
&=\beta_0^+ - \beta_0 + \beta_0^- - \beta_0 - \alpha_0\beta_0(\beta_0^++\beta_0^-) \\ \nn
&\quad - \tilde \beta(t)\big(2 + (\alpha_0 + \tilde \alpha(t))(\tilde \beta^+(t) + \beta_0^+ + \tilde \beta^-(t) + \beta_0^-) \big)\\ \nn
&\quad - \tilde \alpha(t)\beta_0\big(\tilde \beta^+(t) + \beta_0^+ + \tilde \beta^-(t) + \beta_0^-)\big)\\ 
&\quad- \tilde \beta^+(t)(\alpha_0\beta_0-1) -\tilde \beta^-(t)(\alpha_0\beta_0-1).
\end{align}
The requirement on $w(n)$ implies that the shift operators are continuous with respect to the norm
$\| . \|_{w,p}$ and the same is true for the multiplication operator with a bounded sequence.
Therefore, using the generalized H\"older inequality yields that \eqref{2.15} is a system of inhomogeneous 
linear differential equations in the Banach space with norm $\| . \|_{w,p}$ and has a local solution 
with respect to this norm (see e.g.\ \cite{Deimling1977} for the theory of ordinary differential 
equations in Banach spaces). Since $w(n) \geq 1$, this solution is 
bounded and the corresponding coefficients $(\tilde \alpha, \tilde \beta)$ coincide with the solution 
$(\alpha, \beta)$ of the $\AL$ system~\eqref{ALequ} from Theorem~\ref{thm2.2}. 

Moreover, $(\tilde \alpha(t), \tilde \beta(t))$ are uniformly bounded
for $t \in (-T,T)$, as writing \eqref{2.15} in integral form yields
\begin{align*}
& \|(\tilde \alpha(t), \tilde \beta(t))\|_{w,p} \leq \|(\tilde \alpha(0), \tilde \beta(0))\|_{w,p} 
+ t C \|(\alpha_0, \beta_0)\|_{w,2p} + C \int_0^t  \|(\tilde \alpha(s), \tilde \beta(s))\|_{w,p} ds
\end{align*}
for some constants $C$.
\end{proof}

Example \eqref{example} in the introduction follows if we let $\tilde \alpha_0 = \tilde \beta_0 \equiv 0$
and
\be \nn
\alpha_0(n)= \frac{a}{n^{\delta}}, \quad \beta_0(n)= \frac{b}{n^{\delta}}, 
\quad a, b \in \C, \quad \delta \geq 0,
\ee
for $n>0$,  $\alpha_0(n)=\beta_0(n)=0$ for $n\leq 0$.
Now choose $p=\infty$ with
\be \nn
w(n)=\begin{cases}(1+n)^{\min(\delta, (\delta +1)/2)}, & n>0, \\
1, & n\leq0,
\end{cases} 
\ee
and apply Theorem~\ref{thm2.4}.

Finally, we remark that if a solution $(\alpha(n,t), \beta(n,t))$ vanishes at two consecutive points 
$n=n_0$, $n=n_0 +1$ in an arbitrarily small time intervall $t \in (t_1, t_2)$, then it vanishes identically
for all $(n,t)$ in $\Z \times \R$, see \cite{KruegerTeschl2009}. In particular, 
a compact support of the solution is not preserved.
The corresponding result for the $\AL$ hierarchy is derived in \cite{KruegerTeschl2009} as well.

\section{Extension to the Ablowitz--Ladik hierarchy}
\label{secth}

In this section we show how our results extend to the $\AL$ hierarchy.
The hierarchy can be constructed by generalizing the matrix $V(z)$ in the zero-curvature 
equation~\eqref{zc} to a $2\times2$ matrix $V_{\ul r}(z)$, $\ul r=(r_-,r_+) \in \N_0^2$, with Laurent 
polynomial entries, see \cite[Sec.\ 3.2]{GesztesyHoldenMichorTeschl:2008} or
\cite{GesztesyHoldenMichorTeschl:2007a}. 
Suppose that $U(z)$ and $V_{\ul r}(z)$ satisfy the 
zero-curvature equation 
\be \label{zc1}
U_t+UV_{\ul r} -V_{\ul r}^+ U=0.
\ee
Then the coefficients 
$\{f_{\ell,\pm}\}_{\ell=0,\dots,r_{\pm}-1}$, 
$\{g_{\ell,\pm}\}_{\ell=0,\dots,r_{\pm}}$, and 
$\{h_{\ell,\pm}\}_{\ell=0,\dots,r_{\pm}-1}$ of the Laurent polynomials in the entries of $V_{\ul r}(z)$
are recursively defined by   
\begin{align} \label{AL0+zc}
\begin{split}
g_{0,+} &= \tfrac12 c_{0,+}, \quad f_{0,+} = - c_{0,+}\alpha^+, 
\quad h_{0,+} = c_{0,+}\beta, \\  
g_{\ell+1,+} - g_{\ell+1,+}^- &= \alpha h_{\ell,+}^- + \beta f_{\ell,+}, 
\quad 0 \le \ell \le r_+ -1,\\ 
f_{\ell+1,+}^- &= f_{\ell,+} - \alpha (g_{\ell+1,+} + g_{\ell+1,+}^-), 
\quad 0 \le \ell \le r_+ -2, \\  
h_{\ell+1,+} &= h_{\ell,+}^- + \beta (g_{\ell+1,+} 
+ g_{\ell+1,+}^-), \quad 0 \le \ell \le r_+ -2,  
\end{split}
\end{align}
and  
\begin{align} \label{AL0-zc}
\begin{split}
g_{0,-} &= \tfrac12 c_{0,-}, \quad f_{0,-} = c_{0,-}\alpha, 
\quad h_{0,-} = - c_{0,-}\beta^+, \\ 
g_{\ell+1,-} - g_{\ell+1,-}^- &= \alpha h_{\ell,-} + \beta f_{\ell,-}^-, 
\quad 0 \le \ell \le r_- -1,\\ 
f_{\ell+1,-} &= f_{\ell,-}^- + \alpha (g_{\ell+1,-} + g_{\ell+1,-}^-), 
\quad 0 \le \ell \le r_- -2, \\ 
h_{\ell+1,-}^- &= h_{\ell,-} - \beta (g_{\ell+1,-} + g_{\ell+1,-}^-), 
\quad 0 \le \ell \le r_- -2.
\end{split}
\end{align}
Note that $g_{\ell,\pm}$ are only 
defined up to summation constants $\{c_{\ell,\pm}\}_{\ell=0,\dots,r_\pm}$ 
by the difference equations in \eqref{AL0+zc}, \eqref{AL0-zc}.
In addition, the zero-curvature equation \eqref{zc1} is equivalent to
\[
0 = i\begin{pmatrix}0&\begin{matrix} -i\alpha_t
 - \alpha(g_{r_+,+} + g_{r_-,-}^-) \\
 + f_{r_+ -1,+} - f_{r_- -1,-}^-\end{matrix}\\[2mm]
\begin{matrix}  z\big(-i\beta_t + \beta(g_{r_+,+}^- + g_{r_-,-})\\
 - h_{r_- -1,-} + h_{r_+ -1,+}^-\big)\end{matrix} &0      \end{pmatrix}.
\]
Varying $\ul r \in \N_0^2$, the collection of evolution equations 
\begin{align}   \label{AL_p}
\begin{split}
& \AL_{\ul r} (\alpha, \beta) =
\begin{pmatrix}-i\alpha_{t} 
- \alpha(g_{r_+,+} + g_{r_-,-}^-) + f_{r_+ -1,+} - f_{r_- -1,-}^-\\
  -i\beta_{t}+ \beta(g_{r_+,+}^- + g_{r_-,-}) - h_{r_- -1,-} + h_{r_+ -1,+}^- \end{pmatrix}=0,  \\
& \hspace*{7.15cm} t\in\R, \; \ul r=(r_-,r_+) \in\N_0^2,
\end{split}
\end{align}
then defines the time-dependent Ablowitz--Ladik hierarchy.  Explicitly, taking $r_-=r_+$ for simplicity, the first 
few equations are 
\begin{align} \nn
& \AL_{(0,0)} (\alpha, \beta) =  \begin{pmatrix} -i \alpha_t- c_{(0,0)}\alpha \\ 
-i\beta_t+c_{(0,0)}\beta \end{pmatrix} 
=0,\\ \nn
& \AL_{(1,1)} (\alpha, \beta) =  \begin{pmatrix}  
-i \alpha_t- \gamma (c_{0,-}\alpha^- + c_{0,+}\alpha^+) 
- c_{(1,1)} \alpha \\
-i\beta_t+ \gamma (c_{0,+}\beta^- + c_{0,-}\beta^+) +
c_{(1,1)} \beta\end{pmatrix}=0,\\  \nn
& \AL_{(2,2)} (\alpha, \beta) =  \begin{pmatrix}\begin{matrix}-i \alpha_t-
\gamma \big(c_{0,+}\alpha^{++} \gamma^+ + c_{0,-}\alpha^{--} \gamma^-\\
- \alpha (c_{0,+}\alpha^+\beta^- + c_{0,-}\alpha^-\beta^+)
- \beta (c_{0,-}(\alpha^-)^2 + c_{0,+}(\alpha^+)^2)\big)\end{matrix}\\[3mm] 
\begin{matrix}-i\beta_t+
 \gamma \big(c_{0,-}\beta^{++} \gamma^+ + c_{0,+}\beta^{--} \gamma^-\\
- \beta (c_{0,+}\alpha^+\beta^- + c_{0,-}\alpha^-\beta^+)
- \alpha (c_{0,+}(\beta^-)^2 + c_{0,-}(\beta^+)^2)\big)\end{matrix}\end{pmatrix}  \nn \\ 
 & \hspace{3cm} + \begin{pmatrix}
-\gamma (c_{1,-} \alpha^- + c_{1,+} \alpha^+) - c_{(2,2)} \alpha\\
 \gamma (c_{1,+} \beta^- + c_{1,-} \beta^+) + c_{(2,2)} \beta\end{pmatrix}
=0, \, \text{ etc.,} 
\end{align}
where we abbreviated $c_{\ul r} = (c_{r,-} + c_{r,+})/2$ and $\gamma = 1 - \alpha \beta$.
Different ratios of $c_{0,+}/c_{0,-}$ lead to different hierarchies.
The $\AL$ system~\eqref{ALequ} corresponds to the case $\ul r=(1,1)$, $c_{0,\pm}=1$, and $c_{(1,1)}=-2$.
The special choices $\beta=\pm \ol \alpha$, $c_{0,\pm}=1$ lead to the discrete NLS hierarchy,
the choices $\beta=\ol \alpha$, $c_{0,\pm}=\mp \I$ yield the hierarchy of Schur flows. 
The $\AL$ hierarchy is invariant under the scaling transform 
\be \label{rescaling}
\{(\alpha(n), \beta(n))\}_{n\in \Z} \rightarrow 
\{(c\, \alpha(n), \beta(n)/ c)\}_{n\in \Z}, \quad c \in \C \backslash \{0\}.
\ee
Hence choosing $c=e^{\I c_{\ul r}t}$ it is no restriction to assume $c_{\ul r}=0$.

By \cite{GesztesyHoldenMichorTeschl:2007b}, the $\AL$ hierarchy is equivalent to the Lax equation 
\begin{equation}\label{Lax}
\f{d}{dt} L(t) - [P_{\ul r}(t), L(t)] =0, \quad t\in\R, \quad \ul r\in\N_0^2, 
\end{equation}
where $L$ is the doubly infinite five-diagonal matrix  \eqref{L} and (recall \eqref{L_pm})
\[
P_{\ul r} = \f{i}{2} \sum_{\ell=1}^{r_+} c_{r_+ -\ell,+} \big( (L^\ell)_+ - (L^\ell)_- \big)
- \f{i}{2} \sum_{\ell=1}^{r_-} c_{r_- -\ell,-} \big( (L^{-\ell})_+ - (L^{-\ell})_- \big)  -
 \f{i}{2} c_{\ul r} Q_d.
\]

Since the $\AL$ flows are autonomous and $f_{r_\pm-1,\pm}$, $g_{r_\pm,\pm}$, and $h_{r_\pm-1,\pm}$ depend
polynomially on $\alpha, \beta$ and their shifts, \cite[Thm 4.1.5]{AbrahamMarsdenRatiu1988}
implies local existence, uniqueness, and smoothness of the solution of the initial value problem of the 
hierarchy as well (see \cite[Sec.\ 3.8]{GesztesyHoldenMichorTeschl:2008}, 
\cite{GesztesyHoldenMichorTeschl:2007b}).

\begin{theorem} \lb{thm3.1} 
Let $t_{0}\in\R$ and suppose $\alpha_0, \beta_0 \in \ell^p(\Z)$ for some 
$p\in [1,\infty)\cup\{\infty\}$. Then the $\ul r$th Ablowitz--Ladik initial value problem
\begin{equation}
\AL_{\ul r}(\alpha,\beta)=0, \quad (\alpha,\beta)\big|_{t=t_0} 
= (\alpha_0, \beta_0)  \lb{ALh5.23}
\end{equation}
for some $\ul r\in\N_0^2$, has a unique, local, and smooth solution in time, that is, there exists a $T>0$ such that
$\alpha(\dott), \, \beta(\dott) \in C^\infty((t_0-T,t_0+T),\ell^p(\Z))$.
\end{theorem}

Next we show that short-range perturbations of bounded solutions remain short-range.
In fact, we will be more general to include perturbations of steplike background solutions 
as for example \eqref{steplike}.

\begin{lemma} \label{lemma1}
Let $w(n)\geq 1$ be some weight with $\sup_n(|\frac{w(n+1)}{w(n)}|+|\frac{w(n)}{w(n+1)}|)<\infty$
and fix $1\leq p\leq \infty$. Suppose $(\alpha(t), \beta(t))$ and $(\alpha_{\ell,r}(t), \beta_{\ell,r}(t))$ 
are arbitrary bounded solutions of some equation $\AL_{\ul r}$ in the $\AL$ hierarchy and abbreviate 
\be
\tilde \alpha(n,t)= \begin{cases} \alpha_r(n,t), & n\geq 0,\\ \alpha_\ell(n,t), & n<0,\end{cases}\qquad
\tilde \beta(n,t)= \begin{cases} \beta_r(n,t), & n\geq 0,\\ \beta_\ell(n,t), & n<0.\end{cases}
\ee
If
\be \label{H t2}
\|(\alpha(t)-\tilde \alpha(t),\beta(t)-\tilde \beta(t))\|_{w,p}
< \infty
\ee
holds for one $t=t_0 \in \R$, then it holds for all $t \in (t_0-T,t_0+T)$.
\end{lemma}

\begin{proof}
Without loss we assume that $t_0 = 0$. First we derive the differential
equation for the differences 
$\delta(n,t)= \big(\alpha(n,t) - \tilde \alpha(n,t), \beta(n,t) - \tilde \beta(n,t) \big)$
in the Banach space of pairs of bounded sequences $\delta=(\delta_1,\delta_2)$ for which
the norm $\| \delta \|_{w,p}$ is finite.

Let us show by induction on $r_\pm$ that $f_{r_\pm-1, \pm}(t) - \tilde f_{r_\pm-1, \pm}(t)$,  
$g_{r_\pm, \pm}(t) - \tilde g_{r_\pm, \pm}(t)$, and 
$h_{r_\pm-1, \pm}(t) - \tilde h_{r_\pm-1, \pm}(t)$
can be written as a linear combination of shifts of $\delta$ with the coefficients depending only on 
$(\alpha(t),\beta(t))$ and $(\alpha_{\ell,r}(t), \beta_{\ell,r}(t))$. 
It suffices to consider the homogeneous case where 
$c_{j,\pm}=0$, $1\leq j \leq r_\pm$, since all involved sums are finite. 
In this case \cite[Lemma~A.3]{GesztesyHoldenMichorTeschl:2007a} yields that 
$f_{j, +}$, $g_{j, +}$, and $h_{j, +}$ can be recursively computed from 
$f_{0, +}=- \alpha^+$, $g_{0, +}= \frac{1}{2}$, and $h_{0, +}=\beta$ 
via
\begin{align*}
f_{\ell+1,+}^- &= f_{\ell,+} - \alpha (g_{\ell+1,+} + g_{\ell+1,+}^-),\\
h_{\ell+1,+} &= h_{\ell,+}^- + \beta (g_{\ell+1,+} + g_{\ell+1,+}^-),\\
g_{\ell+1,+} &= \sum_{k=0}^\ell f_{\ell-k,+} h_{k,+}
- \sum_{k=1}^\ell g_{\ell+1-k,+} g_{k,+},
\end{align*}
and similarly for the minus sign and $\tilde f_{j,\pm}$, $\tilde g_{j,\pm}$, and $\tilde h_{j,\pm}$.
The fact that 
$(\tilde \alpha, \tilde \beta)$ does not solve $\AL_{\ul r}$ only affects finitely many terms and gives 
rise to an inhomogeneous term $B_{\ul r}(t)$ which is nonzero only for a finite number of terms.

Hence $\delta$ satisfies an inhomogeneous linear differential equation of the form
\[
\I \frac{d}{dt}\delta(t) = \sum_{|j|\leq \max(r_-,r_+)} A_{\ul r,j}(t) (S^+)^j \delta(t) + B_{\ul r}(t)
\]
Here $S^\pm(\delta_1(n,t),\delta_2(n,t))=(\delta_1(n\pm 1,t),\delta_2(n\pm 1,t))$ are the shift operators,
\[
A_{\ul r,j}(n,t)= \begin{pmatrix}
A_{\ul r,j}^{11}(n,t) & A_{\ul r,j}^{12}(n,t)\\ A_{\ul r,j}^{21}(n,t) & A_{\ul r,j}^{22}(n,t)
\end{pmatrix},
\]
are multiplication operators with bounded $2\times 2$ matrix-valued sequences,
and $B_{\ul r}(n,t)= \big(B_{\ul r,1}(n,t), B_{\ul r,2}(n,t)\big)$
with $B_{r,i}(n,t)=0$ for $|n|>\max(r_-,r_+)$.
All entries of $A_{\ul r,j}(t)$ and $B_{\ul r}(t)$ are polynomials with respect to
$(\alpha(n+j,t),\beta(n+j,t))$, $(\alpha_{\ell,r}(n+j,t),\beta_{\ell,r}(n+j,t))$, 
$|j|\leq \max(r_-,r_+)$. 
Thus $\|B_{\ul r}(t)\|_{w,p} \le D_{\ul r}$, where the constant depends only on 
the sup norms of $(\alpha(t),\beta(t))$ and $(\alpha_{\ell,r}(t),\beta_{\ell,r}(t))$.
Moreover, by our assumption the shift operators are continuous,
\[
\|S^\pm\| = \begin{cases} \sup_{n\in\Z} |\frac{w(n)}{w(n\pm1)}|^{1/p}, & p\in[1,\infty),\\
 \sup_{n\in\Z} |\frac{w(n)}{w(n\pm1)}|, & p=\infty, \end{cases}
\]
and the same is true for the multiplication operators $A_{\ul r,j}(t)$ whose norms depend only on the supremum
of the entries by H\"older's inequality, that is, again on the sup norms of $(\alpha(t),\beta(t))$ and
$(\alpha_{\ell,r}(t),\beta_{\ell,r}(t))$.
Consequently, for $t\in (-T,T)$ there is a constant such that 
$\sum_{|j|\leq \max(r_-,r_+)} \|A_{\ul r,j}(t)\| \|(S^+)^j\| \le C_{\ul r}$.
Hence
\[
\|\delta(t)\|_{w,p} \le \|\delta(0)\|_{w,p} + \int_0^t \big(C_{\ul r} \|\delta(s)\|_{w,p} + D_{\ul r}\big)
\]
and Gronwall's inequality implies
\[
\|\delta(t)\|_{w,p} \le \|\delta(0)\|_{w,p} \E^{C_{\ul r} t} + \frac{D_{\ul r}}{C_{\ul r}} \left( \E^{C_{\ul r} t} -1\right).
\]
Since $w(n)\ge 1$, this solution is again bounded and hence coincides with the solution of the 
$\AL$ equation from Theorem~\ref{thm3.1}.
\end{proof}

For certain equations in the $\AL$ hierarchy, i.e.\ for certain configurations of summation coefficients 
$\{c_{j,\pm}\}$, our main result remains valid.

\begin{theorem}
Let $\ul r=(r_-, r_+) \in \N_0^2 \backslash (0,0)$ and assume that $c_{j,\pm} \in \C$, $j=0, \dots, r_\pm$, 
satisfy 
\be \label{constraint}
\sum_{j=0}^{r_+-1} c_{j,+} + \sum_{j=0}^{r_--1} c_{j,-}=0.
\ee
Let $w(n)\geq 1$ be some weight with $\sup_n(|\frac{w(n+1)}{w(n)}|+|\frac{w(n)}{w(n+1)}|)<\infty$.
Fix $1 \leq p \leq \infty$ and suppose $\alpha_0$, $\beta_0$ and $\tilde \alpha_0$, $\tilde \beta_0$ 
are bounded sequences such that
\begin{align*} \label{decay}
& \begin{array}{ll}
 \|(\alpha_0, \beta_0)\|_{w,2p} < \infty, &  \|(\alpha_0-\alpha_0^+, \beta_0-\beta_0^+)\|_{w,p} < \infty, \\
 \|(\tilde \alpha_0, \tilde \beta_0)\|_{w,p} < \infty, & \\
\end{array}\qquad \text{if $1 \leq p < \infty$,} \\[2mm]
& \begin{array}{ll}
 \|(\alpha_0, \beta_0)\|_{w,\infty} < \infty, &
\|(\alpha_0-\alpha_0^+, \beta_0-\beta_0^+)\|_{w^2,\infty} < \infty, \\
 \|(\tilde \alpha_0, \tilde \beta_0)\|_{w^2,\infty} < \infty, & 
\end{array} \qquad \text{if $p=\infty$.}
\end{align*}
Let $(\alpha(t),\beta(t))$, $t \in (-T,T)$, be the unique solution of the equation $\AL_{\ul r}(\alpha,\beta)=0$
with summation coefficients $\{c_{j,\pm}\}_{j=0}^{r_\pm}$, corresponding to the initial conditions
\be
\alpha(0) = \alpha_0 + \tilde \alpha_0, \qquad \beta(0) = \beta_0 + \tilde \beta_0.
\ee 
Then this solution is of the form
\be
\alpha(t) = \alpha_0 + \tilde \alpha(t), \qquad \beta(t) = \beta_0 + \tilde \beta(t), 
\ee
where $\|(\tilde \alpha(t), \tilde \beta(t))\|_{w,p} < \infty$, respectively,
$\|(\tilde \alpha(t), \tilde \beta(t))\|_{w^2,\infty} < \infty$.
\end{theorem}

\begin{proof}
The proof is similar to the one of Theorem~\ref{thm2.4}. From $\AL_{\ul r}(\alpha,\beta)=0$ we obtain
an inhomogeneous differential equation for $(\tilde \alpha, \tilde \beta)$. The homogeneous part is a finite
sum over shifts of $(\tilde \alpha, \tilde \beta)$. The inhomogeneous part consists of products of $\alpha_0$, $\beta_0$
and their shifts, whose $\|.\|_{w,p}$ norm is finite by H\"older's inequality, and 
of sums of the form $c_{j,\pm}\alpha_0$, $c_{j,\pm}\beta_0$ and shifts thereof,
\begin{align*}
-\big(&c_{0,+}S^{+r_+} + c_{1,+}S^{+r_+ -1} + \dots + c_{r_+-1,+}S^{+1} + c_{\ul r} \\ 
&+c_{0,-}S^{-r_-} + c_{1,-}S^{-r_- -1} + \dots + c_{r_--1,-}S^{-1}\big)\alpha_0,
\end{align*}
(and analogously for $\beta_0$) from which restriction \eqref{constraint} arises. 
Again $S^{\pm j}$ denote the shift operators $S^{\pm j}\alpha_0(n)=\alpha_0(n\pm j)$. 
The requirement $\|(\alpha_0-\alpha_0^+, \beta_0-\beta_0^+)\|_{w,p} < \infty$ 
yields the algebraic constraint \eqref{constraint} for $c_{j,\pm}$. 
Finally, note that it is no restriction to assume $c_{\ul r} = 0$ by \eqref{rescaling}.
\end{proof}

For example, we obtain such decaying solutions for $\AL_{(0,1)}(\alpha,\beta)$ if $c_{0,+}=0$,
for $\AL_{(1,1)}(\alpha,\beta)$ if $c_{0,+}=-c_{0,-}$ (or
$c_{0,+}+c_{0,-}+c_{\ul 1}=0$ as in Theorem~\ref{thm2.4}).\\[2mm]
\noindent {\bf Acknowledgment.} The author thanks G.\ Teschl for valuable discussions on this topic.

\end{document}